# Time Series Analysis and Forecasting of the US Housing Starts using Econometric and Machine Learning Models

by Sudiksha Joshi, Mount Holyoke College, B.A, Class of 2019

## Abstract

In this research paper, I have performed time series analysis and forecasted the monthly value of housing starts for the year 2019 using several econometric methods - ARIMA(X), VARX, (G)ARCH and machine learning algorithms - artificial neural networks, ridge regression, K-Nearest Neighbors, and support vector regression, and created an ensemble model. The ensemble model stacks the predictions from various individual models, and gives a weighted average of all predictions. The analyses suggest that the ensemble model has performed the best among all the models as the prediction errors are the lowest, while the econometric models have higher error rates.

## 1. Introduction[1]

In this empirical paper, I have forecasted the housing starts in the United States. Housing starts are the number of new residential construction projects that have begun in a month. The common techniques of forecasting include econometric time series modeling and machine learning (ML). I have chosen housing starts as it is a leading indicator in the real estate or mortgage market . This forward looking variable estimates a good gauge for future levels of real estate supply, and creates a ripple effect in the overall economy. Its is primarily of interest as the inception and collapse of the housing bubble in 2007-08 were the turning points in the subsequent developments that embroiled the American and the Eurozone economies in a deep-seated financial meltdown. Buying new houses also increases the demand of complementary durable goods such as furniture, refrigerators, etc. Thus, new residential construction boosts employment in construction, raw materials, banking and other manufacturing sectors. Mortgage rates directly affect housing activity as higher interest rates raise the housing expenses. This lowers the number of qualified borrowers, declining home sales, and housing starts. Contrarily, lower interest rates make houses affordable, spurring housing starts and home sales. The next section describes the data and methods used.

---

[1] I would like to thank Professor Evan Ray, Professor Michael Robinson, Abena Danso and Marihah Idroos. All errors are my responsibility.

## 2. Data Analysis

### 2.1. Description of the Data

I have used monthly data of all variables from Federal Reserve Bank of St. Louis Economic Research (FRED) from June 1976 to December 2018. The variables are:

1. **hous_st :** The number of privately owned housing starts for each month (in 1000s of units).
2. **CPI:** The consumer price index measures the weighted average of prices of a fixed basket of consumer goods and services. It assesses changes in cost of living standards with respect to inflation. A higher CPI implies higher inflation, reducing people's disposable income/savings, and vice versa.
3. **mortgR:** A 30 year mortgage rate that measures the interest rate charged when financing a new home.
4. **fed_fundsR:** The federal funds rate is the interest rate that banks charge other other banks for overnight loan requests to meet the federal reserve requirement.
5. **income:** The real disposable personal income per person on average (in billions of USD).
6. **pvt_house_comp:** The number of new privately owned houses completed (in 1000s of units).
7. **sec_conL:** The number of securitized total consumer loans outstanding (in billions of USD). Through securitization, financial institutions distribute various assets such as residential mortgages, and auto loans to third party investors to generate a sustainable cash flow.
8. **real_estL:** The number of real estate loans from all commercial banks in billions of USD.
9. **yield_sp:** The difference between a 10 year treasury bond and a 2 year treasury bond.
10. **unempR:** The civilian unemployment rate.
11. **house_supply:** Monthly Supply of Houses in the United States, Months' Supply

I have divided the original dataset into the training and test sets, wherein the first 80 percent of observations belong to the training set, and the last 20 percent belong in the test set. The mean absolute percent error and percent bias measure each model's performance. The mean absolute percentage error (MAPE) is the absolute percent difference between the predicted and observed values and is useful when we compare forecasts of series in different scales. Here, we measure the error in terms of percentage.

$$MAPE = \frac{1}{n} \sum_{i=1}^{n} \frac{|y_i - \hat{y}_i|}{|y_i|}$$

The percent bias calculates the the average amount by which the observed values deviates from the predicted values. Its value is close to 0 if the model is unbiased. From the bias-variance tradeoff, complex models have lower values of bias and higher variance, and vice-versa.

$$PB = \frac{1}{n} \sum_{i=1}^{n} \frac{y_i - \hat{y_i}}{y_i} \times 100\%$$

To ensure that all the variables in the time series are stationary, the ACF and PACF plots below check if the series is autocorrelated.

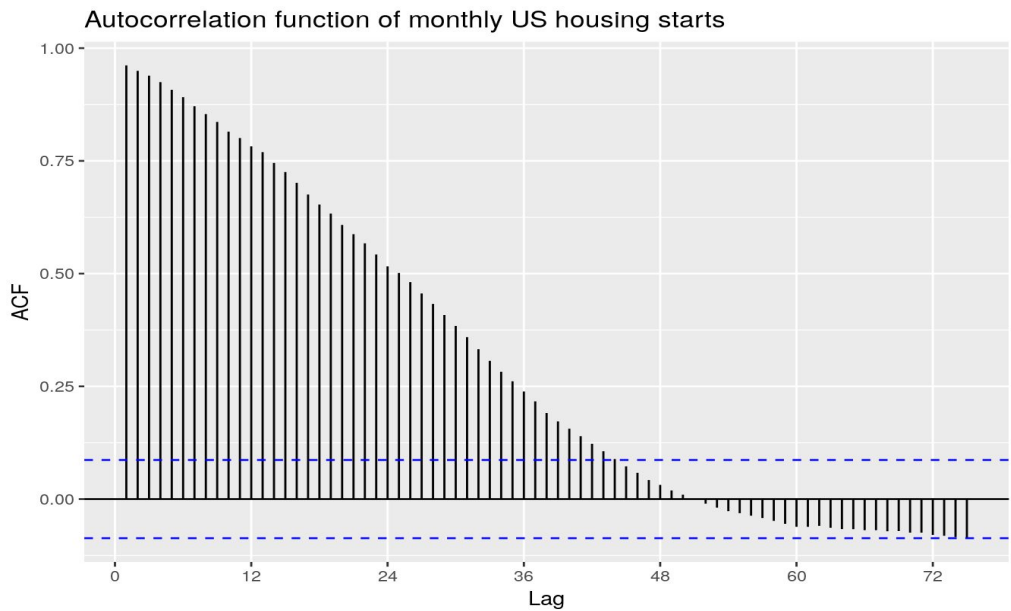

*Figure 1: Plot of the Autocorrelation Function (ACF) of the US Housing Starts shows serial correlation as the ACF dies down*

Because the housing starts series has autocorrelation, it is not white noise. From figure 2, we will include 3 lags. The number of significant correlations (spikes greater than than the significance lines) indicate the term *p* of the AR model.

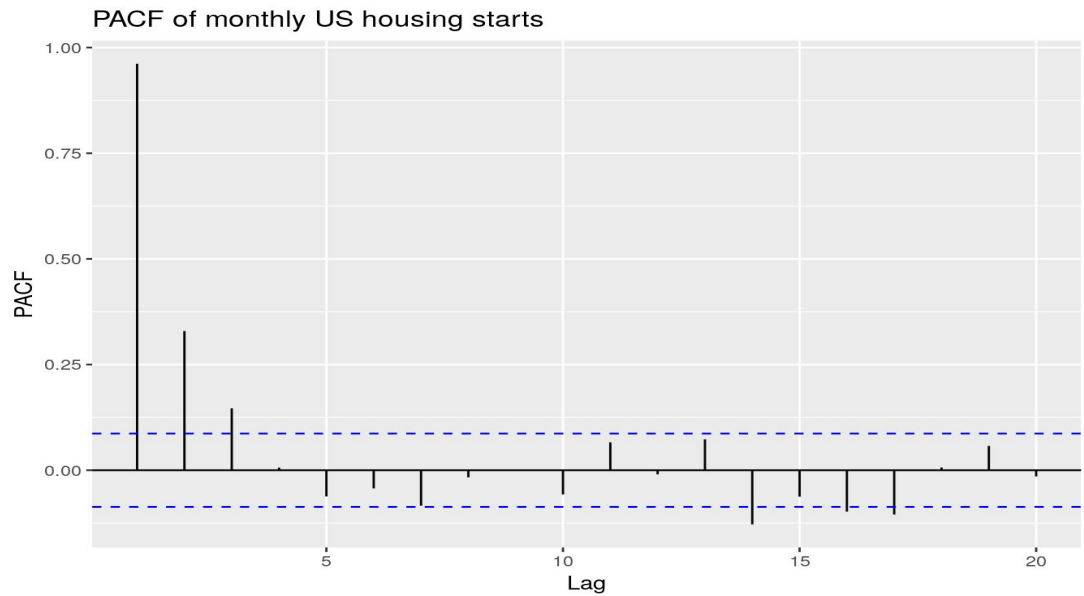

*Figure 2: Partial Autocorrelation Function (ACF) of the US Housing Starts as PACF cuts off after 3 lags*

The ACF of the differenced housing starts doesn't look like that of a white noise series. First differencing removed the trend in the residuals, resulting in uncorrelated errors. There are autocorrelations lying outside the 95% limits, and the Ljung-Box Q∗ statistic has a very small p-value of $3.892 \times 10^{-11}$. This suggests that the monthly values of the US housing starts are not random, but are correlated with those of the previous months. As the Augmented Dickey Fuller (ADF) test concluded that the predictors are non-stationary, I have differenced each of them to make them stationary.

| Variable | P-value before differencing | Number of differences |
| --- | --- | --- |
| Housing Starts | 0.3790981 | 1 |
| Income | 0.6345780 | 2 |
| Federal Funds Rate | 0.1145223 | 1 |
| Yield Spread | 0.2567 | 1 |
| Securitized consumer loans | 0.9705642 | 2 |
| Unemployment rate | 0.0859251 | 1 |
| CPI | 0.2545744 | 3 |
| Private houses completed | 0.7298269 | 1 |
| Mortgage rate | 0.0702159 | 1 |
| Real estate loans | 0.6012911 | 2 |
| Housing supply | 0.3389323 | 1 |

*Table 1: The number of differences required for each variable to be stationary*

### 2.1.2. ARIMA(X) Models

Similar to ARIMA, a multivariate regression model is the ARIMAX model wherein the covariates − mortgage rate, first difference of private houses completed, and second difference of income, securitized consumer loans and real estate loans are present on the right hand side of the model. The $ARIMAX(2, 1, 3)$ model follows an $ARMA(2, 3)$, with a first order difference or backshift, $B$ :

$B \, hous \, st_t = hous \, st_t - hous \, st_{t-1}$. This gives a model with covariates at time $t$ and their coefficients:

Regression with $ARIMAX\,(2, 1, 3)$ :

$B \, hous \, st_t = -48.1082 \, mortgR_t - 0.0054 \, income.d2_t + 1.2251 \, sec \, conL.d2_t - 7.1930 \, CPI.d3_t + 0.1023 \, pvt \, house \, comp.d1_t + 0.0096 \, real \, estL.d2_t + \varepsilon_t$, where:

$\varepsilon_t = 1.5159 \, B \, hous \, st_{t-1} - 0.8454 \, B \, hous \, st_{t-2} - 1.9201 \varepsilon_{t-1} + 1.4622 \, \varepsilon_{t-2} - 0.3158 \, \varepsilon_{t-3}$

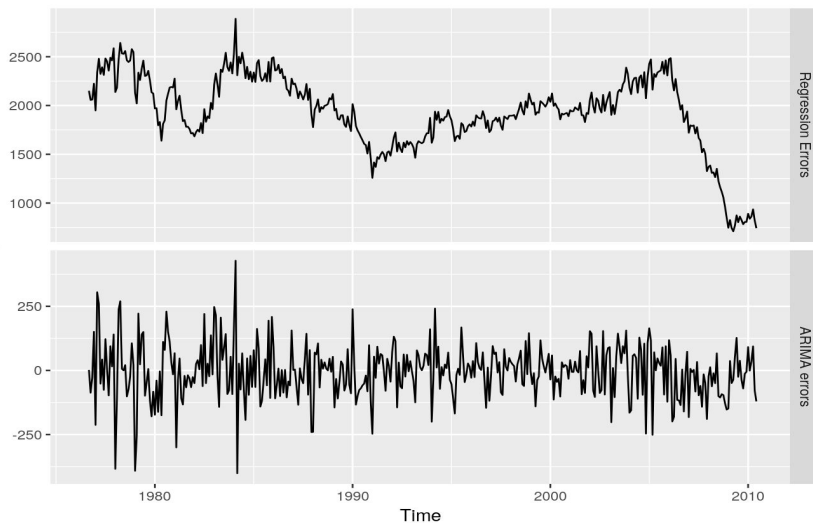

*Figure 3 : The plots show errors from regression and ARIMA models. The ARIMA errors resemble white noise.*

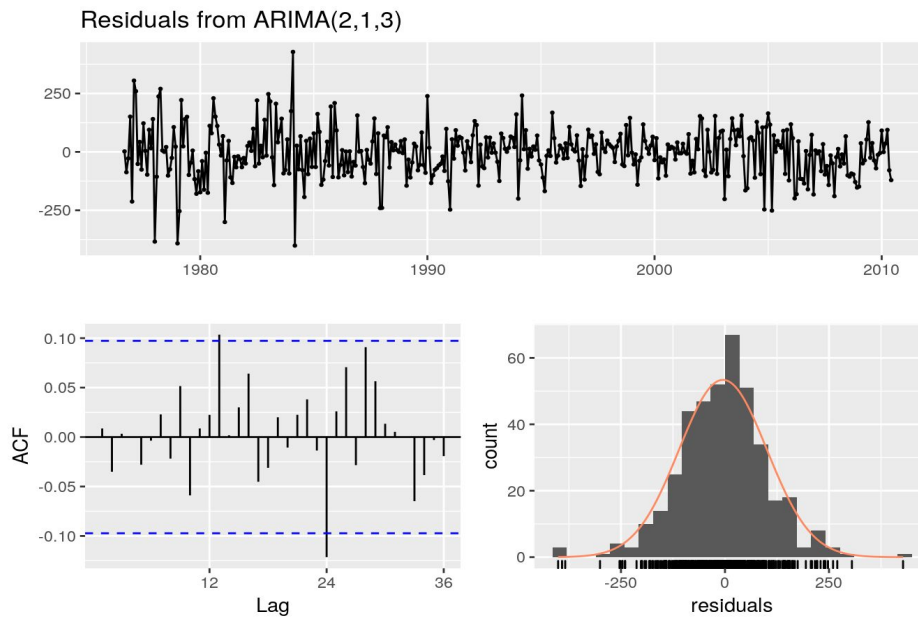

*Figure 4: The diagnostic plots show the ACF and histogram of residuals from ARIMA(2,1,3)*

From the Ljung-Box test on the data comprising the residuals from regression with $ARIMAX(2,1,3)$, the p-value of $0.1069 > 0.05$, implying that the model is not serially correlated. The time plot and histogram of the residuals shows that the variance in the residuals are almost constant. The MAPE is 0.357 and the percent bias is 0.3439.

Unlike the ARIMAX model which contains exogenous variables, the $ARIMA(3,1,2)$ predicts the endogenous variable − housing starts using only its lags and the model is:[2]

$B\ hous\ st_t = 1.0488\ B\ hous\ st_{t-1} - 0.4336\ B\ hous\ st_{t-2} - 0.1437\ B\ hous\ st_{t-3} - 1.427\ \varepsilon_{t-1} + 0.8473\ \varepsilon_{t-2} + \varepsilon_t,$

where $\varepsilon_t$ is white noise. The mean absolute percentage error (MAPE) is 0.3922 and the percent bias is 0.3885. Both the error rates are higher than those of $ARIMAX(2,1,3)$. Figure 5 below shows the plot of residuals of the model to check if the series is discrete white noise (DWN):

---

[2] In R, I had looped over several combinations of p,d and q and stored the fitted model of ARIMA(p,d,q). Given any order of ARIMA, if the current AIC value is less than the previously generated AIC, then the current AIC is the final AIC and that order is chosen. After terminating the loop, the order obtained was ARIMA(3,1,2).

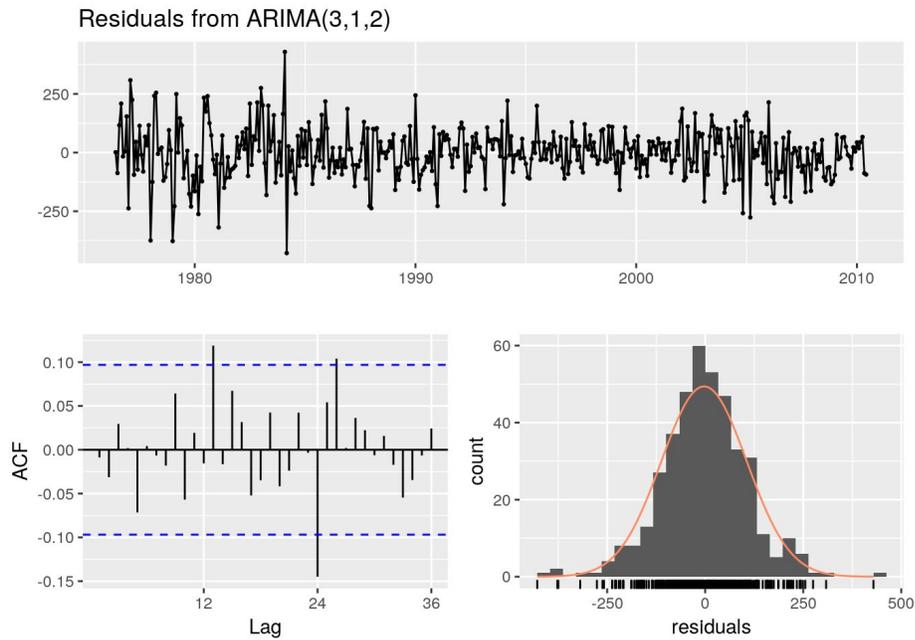

*Figure 5: The ACF plot of the residuals of ARIMA(3,1,2) shows that the errors are not serially correlated.*

The correlogram seems to follow discrete white noise. The p-value of $0.07732 > 0.05$, fails to reject the null hypothesis that the residuals of $ARIMA(3, 1, 2)$ is not serially correlated i.e. the series is white noise.

2.1.3. ARCH and GARCH models

The two main drawbacks of ARIMA models are − firstly, they do not consider volatility clustering i.e. they are not conditionally heteroskedastic. Secondly, because ARIMA linearly models the data, the forecast width is constant as the model does not incorporate new information or recent changes. Hence, we need to use the Autoregressive Conditional Heteroskedastic (ARCH) model and Generalised Autoregressive Conditional Heteroskedastic (GARCH) model to show heteroscedastic variances.

A univariate $GARCH(1, 1)$ helps in modeling volatility, its clustering and forecasting. The PACF and ACF of squared residuals in figure 6 have no significant lags. Finally, we cannot predict a strict white noise series, either linearly or nonlinearly. The squared residuals of $ARIMA(3, 1, 2)$ model show a cluster of volatility or conditional heteroskedasticity as shown from the ACF plots.

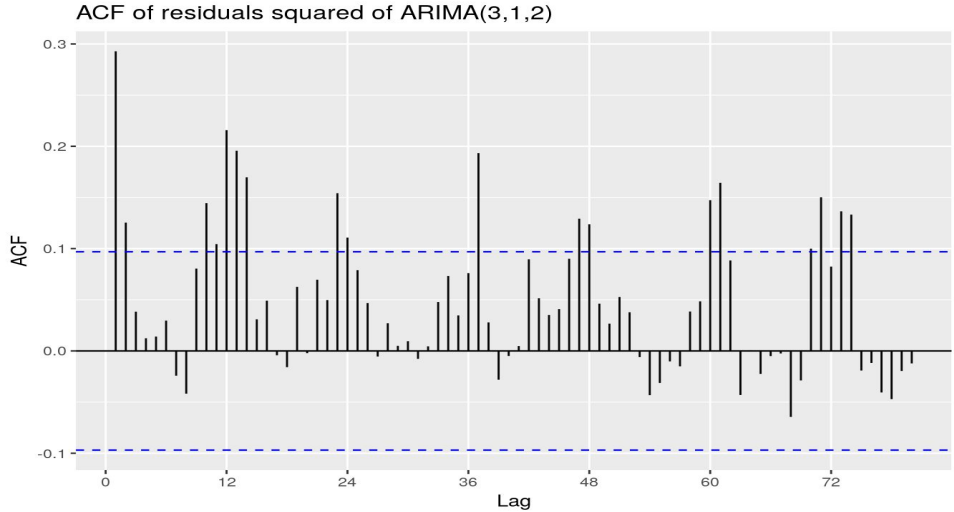

*Figure 6: The ACF plot of the squared residuals of ARIMA(3,1,2) indicate that volatility is clustered together in different time periods.*

The squared residuals from $ARIMA(3, 1, 2)$ are autocorrelated as they do not randomly occur in time, implying that the time series exhibits conditional heteroskedasticity. Also, the conditional heteroskedastic series is non-stationary. This is also called volatility clustering as periods of high variance tend to group up together. The Ljung-Box test on the squared residuals yields a p-value of $2.623 \times 10^8 < 0.05$, which rejects the null hypothesis of no serial correlation. To further prove the presence of conditional heteroskedasticity, I have performed the Engle's ARCH-LM test with $q$ lags. As the p-values with varying values of $q = 2, 4, 8, 12,$ are very small, we reject the null hypothesis to conclude that ARCH effects are present.

An extension of ARCH models, I have fit GARCH model(s), starting with a $GARCH(1, 1)$ model with Gaussian innovations.[3] $GARCH(1, 1)$ considers a single autoregressive and a moving average lag, i.e. $ARMA(1, 1)$. The model predicts the variance by taking the weighted average of the long term historical variance, the predicted variance at time $t$, and the predicted variance of the squared residuals at time $(t-1)$. An $ARMA(1, 1) - GARCH(1, 1)$ model is :

$$y_t = \mu + \theta_1 y_{t-1} + \gamma_1 \varepsilon_{t-1} + \varepsilon_t$$

$$\varepsilon_t = \sigma_t z_t, \text{ where } z_t \sim N(0, 1) \text{ i.i.d., and } \sigma^2_t = \alpha_0 + \alpha_1 \varepsilon^2_{t-1} + \beta_1 \sigma^2_{t-1}, \alpha > 0, \beta \geq 0.$$

---

[3] The gaussian distribution is the same as normal distribution.

$\alpha, \beta$ *and* $z$ are constants that the model estimates using maximum likelihood and $\varepsilon_{t-1}$ is the squared residual at time $(t-1)$. $\theta_n$ and $\gamma_n$ measure $AR(n)$ and $MA(n)$, respectively. In the GARCH model, the persistence explains the rate at which volatility decays after applying a shock. $\alpha_1$ measures the extent to which a volatility shock today feeds through into next period's volatility and $\alpha_1 + \beta_1$ measures the rate at which this effect dies over time. Large values of $\beta_1$ means that large changes in the volatility will affect future volatilities for a long period of time since the decay is slower.

The $ARMA(2,2) - GARCH(1,1)$ model for housing starts is:[4]

$$hous\, st_t = 0.6152\, hous\, st_{t-1} + 0.3685\, hous\, st_{t-2} - 0.4984\, \varepsilon_{t-1} + 0.1791\, \varepsilon_{t-2}$$

$$\sigma^2_t = 2027.8896 + 0.1277\, \varepsilon^2_{t-1} + 0.4095\, \sigma^2_{t-1}, \text{ where } \varepsilon_t \sim iidN(0,1).$$

$AR(1)$ and $AR(2)$ are statistically insignificant, implying that no autocorrelation exists. $\beta_1 = 0.4094$ is also insignificant, which means there is no persistent volatility clustering. To check for normality, we have to observe the distribution of the residuals. The standardized residual tests in Table 2 such as Jarque-Bera Test and the Shapiro-Wilk tests suggest that the data is normally distributed. From the Ljung-Box Test, the distribution of residuals and squared residuals is Gaussian as the p-value is greater than the 5 percent significance level at lags 10, 15 and 20. Similarly, the model has no $ARCH$ effects as the p-value of the $ARCH-LM$ is very high. The table below shows the different residual tests and the p-values:

| **Residual Test** |  | **Statistic** | **p-value** |
|---|---|---|---|
| Jarque-Bera Test | Residual | $\chi^2$ (chi square) | 0.6927309 |
| Shapiro-Wilk Test | Residual | W | 0.9463657 |
| Ljung-Box Test | Residual | $Q(10)$ | 0.1138027 |
| Ljung-Box Test | Residual | $Q(15)$ | 0.05391022 |

---

[4] The order of ARMA - GARCH should be simultaneously determined, not separately. When the ARMA - GARCH models estimate the process well, estimates will be inconsistent if we consider the conditional mean (ARMA) model only and neglect the conditional variance, (GARCH) model as this implicitly assumes that the conditional variance is constant, and vice-versa. Only under certain conditions, such as when MA(0), we can estimate the order of ARMA independent from GARCH and also generate consistent estimates.

Unfortunately, it is difficult to jointly estimate the orders of ARMA - GARCH. The reasonable approach is to experiment with a few candidates models in the training set, pick the model with the lowest prediction rate, and use the selected model to predict in the test set. Ensuring that the prediction rate is lowest does not necessarily guarantee that the model is suitable. We also have to check if the residuals and the squared residuals have ARCH effects, autocorrelation and normal distribution. For instance, ARMA(2,1) - GARCH(1,1) resulted in the lowest MAPE of 0.30, but the residuals failed the diagnostic tests, so I eliminated it.

| Ljung-Box Test | Residual | $Q(20)$ | 0.06944194 |
|---|---|---|---|
| Ljung-Box Test | Residual Squared | $Q(10)$ | 0.2186401 |
| Ljung-Box Test | Residual Squared | $Q(15)$ | 0.1054082 |
| Ljung-Box Test | Residual Squared | $Q(20)$ | 0.1093043 |
| Arch-LM Test | Residual | $TR^2$ | 0.1709148 |

*Table 2: Standardized residual tests and the p-values. All the p-values are significant at 5 percent.*

The qq-plot of the standardised residuals, suggests that the fitted standardised skew-t conditional distribution is good. Using the above model, I have forecasted housing starts for the year 2019 which is shown in section 3: Results.

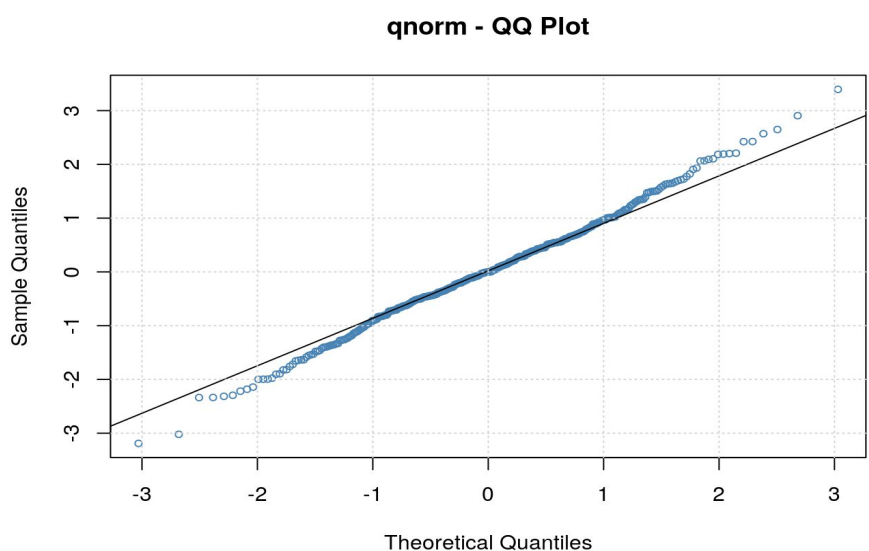

*Figure 7: The plot shows the quantiles of the residuals. Since most of the points lie along the line, the distribution has the same shape as that of the theoretical distribution.*

Now, I will discuss some of the advanced methods that necessitates me to check first if the explanatory variables have common stochastic trends, or are cointegrated.

2.1.4 Cointegration

When the trends and patterns of two series are similar, then they are cointegrated. The cointegration test measures whether the residuals from the regression below are stationary.

$hous\ st_t = 66.8147 + 0.9736\ pvt\ house\ comp_t + r_t,\ \ r_t = 0.7482\ r_{t-1} + \varepsilon_t$

Figure 1 in the Appendix shows the plots of the equation, residuals and innovations. Here, the series seems cointegrated but the residuals are not $AR(1)$. As the p-values (shown below) are very small for all the tests, we reject the null hypothesis that the residuals are unit root, implying the series: housing starts and private houses completed are cointegrated. In fact, no other variable is cointegrated with housing starts.

| Test | p-value |
| --- | --- |
| Augmented Dickey Fuller | 0.00583 |
| Phillips Perron | 0.0001 |
| Schmidt and Phillips Rho | 0.0001 |
| Johansen Trace Test | 0.0001 |

*Table 3: Unit root test for stationarity of residuals from cointegration*

2.1.5. Error Correction Model (ECM)

A drawback of Error-Granger cointegration statistic regression is that it has a small bias, and we cannot infer if the estimated parameters in the regression are significant as the distribution depends upon unknown parameters. To avoid this problem, I have built an error correction model that investigates the long-run relationship between the two series. The variables without cointegration $I(0)$ have a short term relationship as opposed to those variables with cointegration $I(1)$, as the latter have a long term relationship. The ECM is:

$\Delta\ hous\ st_t = -0.25682\ \varepsilon_{t-1} + 0.24530\ \Delta\ pvt\ house\ comp_t + u_t$

From the cointegration part and the error correction term, $\beta = 0.9736$ is the long run parameter, while $\alpha = -0.25682$ and $\gamma = 0.24530$ are the short-run parameters. So far, I have only used one explanatory variable, but to find the number of cointegrating relations when there are more than two $I(1)$ series, we use the Johansen Test and a Vector Autoregressive Model (VAR).

2.1.6. The Johansen Test for Cointegration

This test measures the number of cointegrating relations present out of $n$ integrated time series. Cointegrated series have at least one common trend among the variables. Then we test if the linear combination of underlying series forms a stationary series. If the series $x_1, ..., x_n$ at time $t$ are individually integrated, $I(1)$, then the linear combination, $y = \beta_1 x_1 + \beta_2 x_2 + ... + \beta_n x_n$ at time $t$ is stationary: $I(0)$, and the variables are cointegrated. In this case, I have tested cointegration between variables housing starts, private houses completed, housing supply and real estate loans. The largest eigenvalue generated by the test is $0.1506$. Next, the output shows the trace test statistic for the four hypotheses of $r = 0$ to $r \leq 3$. From $r = 0$ to $r \leq 2$, the test statistic exceeds the $0.05$ significance level. For instance, when $r = 0$, $133.69 > 48.28$. Similarly, in the second test we test the null hypothesis for $r \leq 1$ against the alternative hypothesis of $r > 1$. As $50.59 > 31.52$, we reject $r \leq 1$, i.e. the null hypothesis of no cointegration. However, when $r \leq 2$, we fail to reject the null as $12.87 < 17.95$. Thus, the matrix' rank is 2 and the series will become stationary after using a linear combination of two time series.

| $r$ = rank | Test statistic | 10 % level | 5 % level | 1 % level |
|---|---|---|---|---|
| $r \leq 3$ | 0.65 | 6.50 | 8.18 | 11.65 |
| $r \leq 2$ | 12.87 | 15.66 | 17.95 | 23.52 |
| $r \leq 1$ | 50.59 | 28.71 | 31.52 | 37.22 |
| $r = 0$ | 133.69 | 45.23 | 48.28 | 55.43 |

*Table 4: Values of trace statistics at significance levels for the hypotheses*

The linear combination by using components of eigenvectors associated with the largest eigenvalue of $0.1506$ results in the following stationary series:

$linear\ series\ =\ 1.00\ house\ st\ -0.9398363\ pvt\ house\ comp$.

Moreover, the p-value in the Dickey-Fuller test is 0.01 < 0.05. So, we reject the null hypothesis of unit root and conclude that the series formed from the linear combination is stationary.

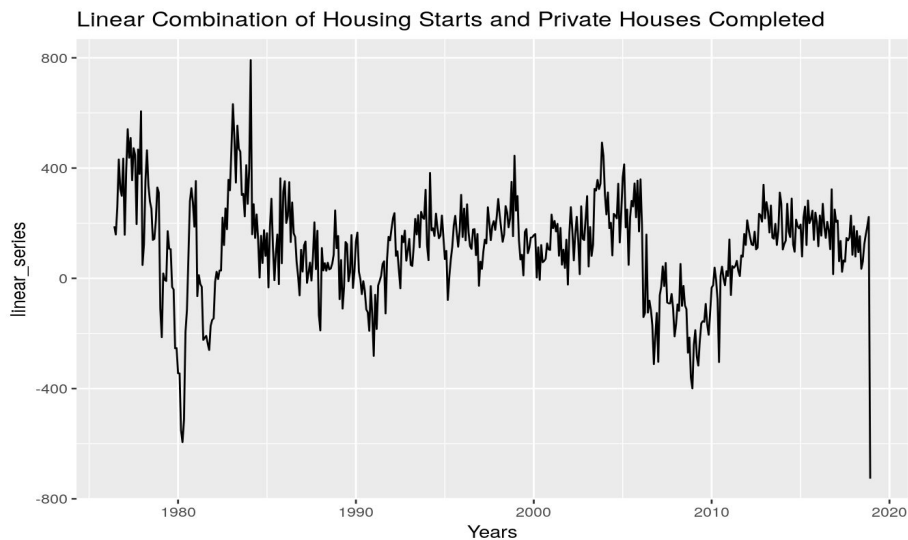

*Figure 8: The plot shows the linear combination of two series as there are two cointegrating relations.*

Having obtained a linear combination of cointegrated series, I have made a Vector Error Correction Model to model the cointegrated time series.

2.1.7. Vector Error Correction Model (VECM)

Unlike the error correction model which is a single equation, a VECM is a multiple equation model used when series are non-stationary and cointegrated. Using the same variables used to check for cointegration, I made a VECM model with 3 lags and 2 cointegrating relations (check Table 1 in the Appendix for the full model). The value of lag which results in the minimum information criteria (in this case, AIC) is chosen as the order of the model. Table 5 depicts only the error correction terms (ECT1 and ECT2) and the p-values of the the four variables. As the p-values are extremely small, the coefficient estimates are significant at the 5 percent critical value.

| Variables | ECT 1 | ECT 2 | p-value |
| --- | --- | --- | --- |
| Housing Starts | -0.0267838950 | -0.0363147859 | 5.283760e-01 <br> 3.335254e-01 |
| Private Houses | 0.3198985285 | 0.3288063833 | 8.922016e-19 |
| Completed | | | 3.075843e-24 |
| Mortgage Rate | 0.0002606689 | -0.0001891956 | 1.870030e-02 |
| | | | 5.323680e-02 |
| Housing Supply | 0.0003273892 | -0.0002339375 | 1.274370e-01 |
| | | | 2.175834e-01 |

*Table 5: Coefficient estimates and p-values of the two error correction terms of the four variables*

The error correction terms (ECT) describe how the time-series adjust to disequilibrium and ECTs are between 0 and 1. The ECTs for housing starts are negative. This implies that when $hous\ st_{t-1}$ is above its long-run level, then there will be a negative change in $\Delta\ hous\ st_t$, which would pull housing starts back towards its long-run relationship with mortgage rate, housing supply and private houses completed. The estimated coefficient $ECT(-1)$ of $hous\ st$ is -0.0268, suggesting that 2.68 percent of disequilibrium is corrected between between one month. The coefficients of ECT1 and ECT2 of housing supply and mortgage rate are almost negligible, denoting that there are dominant variables and do not adjust towards the equilibrium.

While VECM tests for long term relationship, a VARX captures short-run relationship among the variables employed (example, where there is a shock). We can transform a VECM into a Vector Autoregression Model (VARX) to forecast values.

2.1.8. Vector Autoregression with Exogenous Variables (VARX)

The VARX model can be used when the variables under study are $I(1)$ but not necessarily cointegrated. I estimated a $VARX(3)$ model (Table 2 in the Appendix) with with four predictors − housing starts, housing supply, mortgage rate and private houses completed. The appropriate $p$ number of lags is selected using the usual goodness of fit criteria and then checked if the residuals correspond to the model assumptions− absence of serial correlation, homoscedastic and normally distributed residuals. The lag parameter, $p = 3$ minimizes the information criterion: AIC. The large p-value of $0.0786 > 0.05$ suggests that the residuals for this model are not serially correlated, so the model is appropriate. [5] I have forecasted the values of housing starts using $VARX(3)$ in the results section. But "a good forecast does more than provide a current 'best estimate'—it identifies the key variables and states the nature of their impact. As new data come in, such a forecast gives a basis for continual reappraisal of the situation." [6] So, I have performed structural analyses such as impulse response functions and granger causality test that summarize the properties of a $VARX(p)$.

---
[5] Difference between VARX and ARIMAX: VARX is devoid of MA terms and uses autoregressive lags to approximate MA terms. It's solution is a less parsimonious solution than when we directly include MA terms in the ARIMAX model. OLS or GLS can estimate VARX quickly, whereas maximum likelihood method estimates ARIMAX model, which is usually slow.

2.1.9. Impulse Response Function

In IRF, we shock one variable and propagate it through the fitted VARX model for a number of periods. We can trace this through the VARX model and see if it impacts the other variables in a statistically significant way.

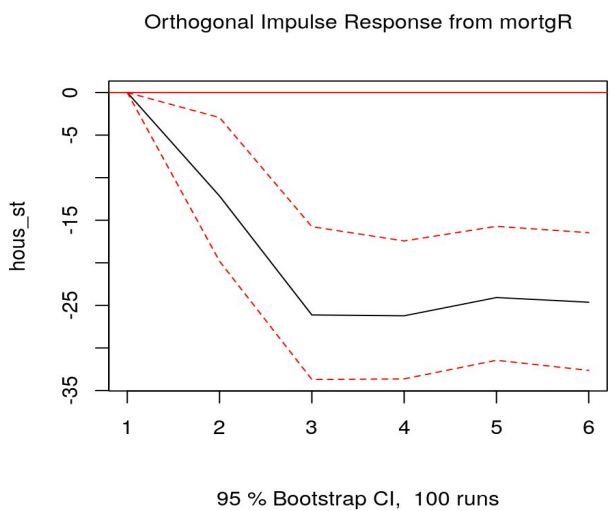

*Figure 9: Impulse response from mortgage rate on housing starts*

When there is a shock in housing starts by mortgage rate, the value of housing starts traverses negatively at a fixed rate away from the mean, creating a momentum effect. After three periods, it hovers around a constant rate, creating a persistantance effect, and doesn't return to its historical mean. The dotted lines show the 95 percent interval estimates of these effects. The VARX function prints the values corresponding to the impulse response graphs.

---

[6] by George Schultz, A Note on Forecasting, 1963

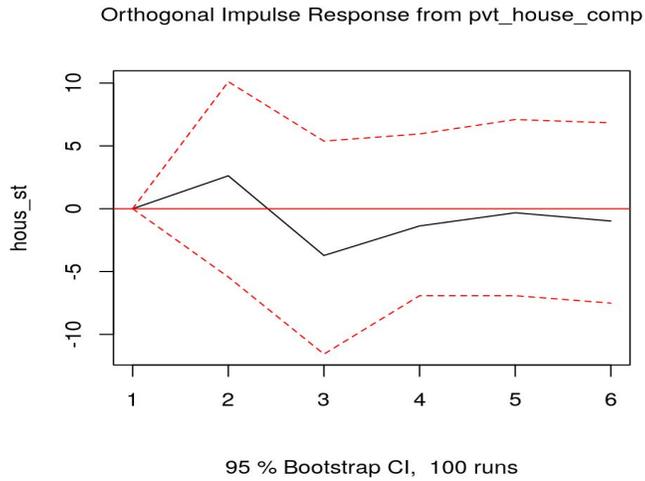

*Figure 10: Impulse response from private houses completed on housing starts*

The impulse response of private houses completed on housing starts shows that with a one unit shock by private houses completed causes housing starts to slightly increases, then fall towards zero into the negative territory at a constant rate as the effect of shock dies, followed by a slight upward movement as housing starts reverts to the mean.

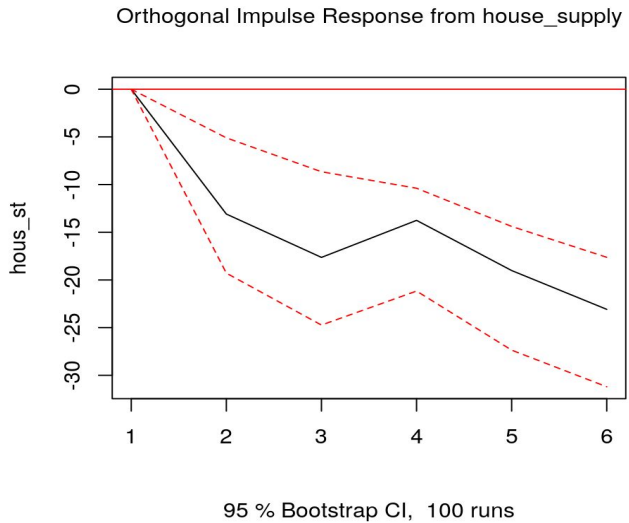

*Figure 11: Impulse response from housing supply on housing starts*

An impulse (shock) to housing starts from housing supply at time zero has large effects the next period as housing starts falls precipitously. The downward momentum persists, and even after six periods of shock, the value of housing starts does not revert to the mean.

2.1.10. Granger Causality

As VARX models forecast each variable, we can describe the relationship between the variables using granger causality test. The F-test on the lags of other variables implements the granger causality. It tests the null hypothesis that all the lags of of a variable $X$ are do not have a predictive power for $Y$, i.e. it does not not contain useful information to predict. In the first case, the F-test rejects the null hypothesis that housing starts do not granger cause private houses completed, mortgage rate and housing separately as the p-value of $1.597 \times 10^{-11} < 0.05$. Likewise, the test rejects the null hypothesis that mortgage rate does not granger cause housing starts, private houses completed and housing supply as the p-value of $2.065 \times 10^{-10}$ is very small. However, private houses completed does not granger cause housing starts, mortgage rate and housing supply as the p-value of $0.6742$ is very high.

| **Null hypothesis** | p-value |
|---|---|
| Housing starts do not granger cause private houses completed, mortgage rate and housing supply | 1.597478e-11 |
| Private houses completed do not granger cause housing starts, mortgage rate and housing supply | 0.6742 |
| Mortgage rates do not granger cause private houses completed, housing starts and housing supply | 2.065e-10 |
| Housing supply do not granger cause private houses completed, mortgage rate and housing starts | 4.975e-07 |

*Table 6: p-values of the four different null hypothesis in a Granger Causality Test*

## 2.2. Machine Learning Models

In this section, I have predicted housing starts in the United States using a stacked ensemble of the following machine learning methods: K-Nearest Neighbors, Ridge Regression, Support Vector Regression, and Artificial Neural Networks. Machine learning relies on cross-validation to prevent overfitting and underfitting. Together these aforementioned base learners in the ensemble enhance the predictive accuracy and robustness, which is otherwise not possible from using each of them separately. In setting up the ensemble, I specified the base algorithms used with specific parameters of each model and train each of them in the training set. Then, I conducted an *18*-fold cross validation[7] on each of the learners and obtained the cross-validated predictions from each of them. Models which are more accurate are assigned higher weights in stacked ensemble. The predictive accuracy increases when the ensemble uses diverse set of initial base learners with different hyperparameters and feature subsets. However, such fine-tuning to generate complex models may cause the problem of overfitting where the model perfectly fits the training set, but generalizes poorly in the test set.

### 2.2.2. Principal Component Analysis for Data Preparation

The variables in the whole dataset are highly correlated with each other, causing multicollinearity. These are redundant variables as they add nothing new to the model. As this may bias OLS estimates, I used Principal Components Analysis (PCA). PCA reduces $p-$dimension dataset to an $m$-dimension dataset where $p > m$. It describes the original data using fewer variables or dimensions than initially measured. We project the original data and the differenced variables onto a new, orthogonal basis. This removes multicollinearity.

$PC_i$ reflects as much information from the original variables. If $\Sigma$ is a covariance matrix of the ten explanatory variables $X = \{hous\ st,\ CPI,\ ...,\ house\ supply\}$ with eigenvalues of $\Sigma$: $\lambda_1 \geq \lambda_2 \geq ... \geq \lambda_{10} \geq 0$, and their respective eigenvectors: $e_1,\ e_2,\ ...,\ e_{10}$, then the $i^{th}$ PC is:

$PC_i = e_i^T X = e_{i1} income + e_{i2} fed\ fundsR + e_{i3} CPI + ... + e_{i\ 10}\ house\ supply,\ i = 1, ...10.$

---

[7] Unsure of the best way to evaluate time series models while forecasting, economists and statisticians often evaluate a model's performance in the test set or out of sample (OOS) set. Cross validation does not account for unit roots and serially correlated variables when using econometric models. The former method evaluates in only one set, while the latter (CV method) evaluates in multiple sets. OOS evaluation is a standard procedure as conventional models such as ARIMA are completely iterative i.e. they start estimating from the beginning of the series. Opsomer et. al (2001) explain that if errors are highly autocorrelated, then cross validation "underestimates bandwidths in a kernel estimator regression framework", overfitting the model. Nonetheless, such problems are immaterial when applying ML methods, so we use CV.

$\frac{\lambda_k}{\sum_{i=1}^{10} \lambda_i}$ is the proportion of total information explained by the $k^{th}$ principal component.

In PCA, every variable is centered at zero so that we can easily compare each principal component to the mean. Centering also removes problems arising the scale of each variable. The components are always sorted by how important they are, so the most important components will always be the first few. In this dataset, $PC1$ accounts for more 60 percent of total variance in the data.

| Dimension | Eigenvalue | Percentage of variance | Cumulative Variance |
| --- | --- | --- | --- |
| 1 | 6.0739086 | 60.7390863 | 60.73909 |
| 2 | 1.8835272 | 18.8352721 | 79.57436 |
| 3 | 1.1807085 | 11.8070849 | 91.38144 |
| 4 | 0.4287866 | 4.2878663 | 95.66931 |
| 5 | 0.2441449 | 2.4414486 | 98.11076 |
| 6 | 0.1353575 | 1.3535751 | 99.46433 |
| 7 | 0.0299749 | 0.2997494 | 99.76408 |
| 8 | 0.0182776 | 0.1827762 | 99.94686 |
| 9 | 0.0038175 | 0.0381750 | 99.98503 |
| 10 | 0.0014966 | 0.0149660 | 100.00000 |

*Table 7: Eigenvalues and the variance explained by each principal component*

A common technique to determine the number of PCs to use is to eyeball the scree plot below wherein we observe the "elbow point", where the proportion of variance explained (PVE) plummets. An eigenvalues less than 1 would mean that the component actually explains less than a single explanatory variable. The first 3 components have an eigenvalue greater 1 and explains almost 91 percent of variance. Alternatively, the first 6 components explain 99.46 percent variance. Thus, I have reduced dimensionality from 12 to 6 while only "losing" 0.53567 percent of variance.[8]

---

[8] Albeit the plot suggests that we should use three PCs only as they explain enough variation, I chose six PCs because the MAPE and percent bias in the the test sets were considerably lower when using the latter. So, all the models performed performed better with six PCs.

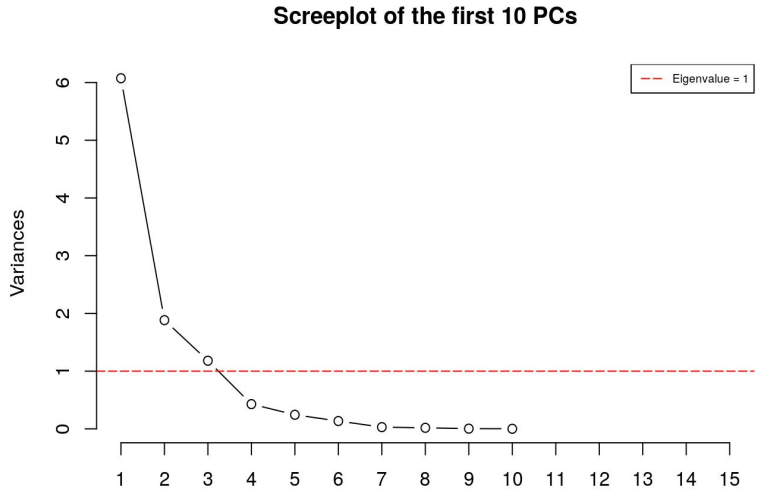

*Figure 12: Visualize eigenvalues (scree plot): shows the percentage of variances explained by each principal component.*

After creating a new dataset with these six principal components and the response variable, housing starts, I split the dataset into training and test − the first 80 percent of the observations are in the training set and the last 20 percent are in the test set. Next, I have explained the individual models used to make the stacked ensemble.

2.2.3. K- Nearest Neighbors

K-Nearest Neighbors (KNN) is one of the simplest machine learning models that is mostly used to classify data points based on how the neighbors are classified (separated into different categories), but also used in regression to predict values. KNN algorithm stores all the available cases and classifies the new case based on how similar it is to the $k$ nearest cases. To find the nearest neighbors, we calculate the Euclidean distance between the new point $(a, b)$ and each point in the training set $(x_i, y_i)$:

$$distance = \sqrt{(a - x_i)^2 + (b - y_i)^2}$$

We chose the $k$ value which determines the number of neighbors to consider before establishing the value to the new observation. The final prediction $f(x_0)$ for the point $x_0$ is the average value of the $k$ training observations $y_i$ that are closest to $x_0$, represented by $N_0$.

$$f(x_0) = \frac{1}{k} \sum_{x \in N_0} y_i$$

In the KNN model with six predictors on the training set, $k$ is set from 0 to 10. Ten different models with all ten values of $k$ are trained and ultimately we select the model with the lowest error. When $k=2$, the error rate is lowest and thereafter, it rises. The test set MAPE of $27.9\%$ and a negative bias of $18.15\%$.

2.2.4. Support Vector Regression

Just as we minimize the error rates in simple regression, in Support Vector Regression, the goal is to fit the error within a particular threshold.

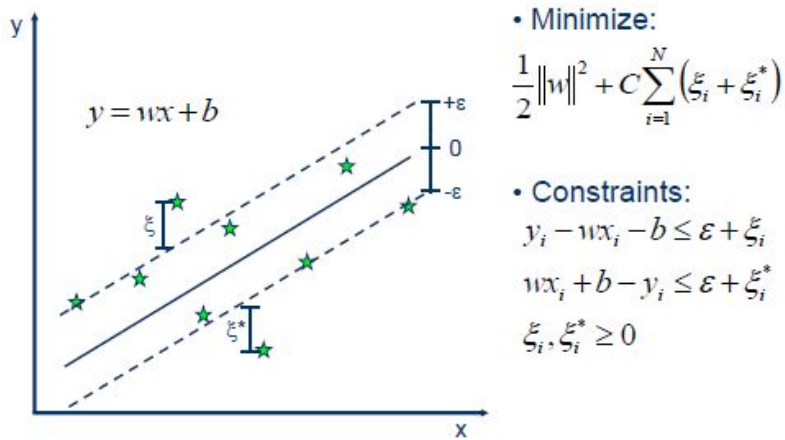

*Figure 13: Hyperplane, boundaries and margin in a support vector regression[9]*

In figure 13, the perpendicular distances between the blue middle line and the observed values (shown by the green stars) nearest to the blue line are called margins. Most of the green stars lie within the black dotted lines (boundary lines). We consider all the points that are within the boundary line when moving along the SVR. The decision boundary is the margin of tolerance. In SVR, we only consider those points that minimize errors, fitting a better model. These stars are known as support vectors, which are closest to the boundary lines. Hyperplane in a two-dimensional space is the blue line in the middle that is useful to predict the target values. In an $n$-dimensional space, it is a $(n-1)$ subspace. The line of best fit is the hyperplane with the maximum number of points. In SVM, the hyperplane separates between data classes.

---

[9] Source: http://www.saedsayad.com/support_vector_machine_reg.htm

The SVR model uses two parameters: the cost and loss values to avoid overfitting. The values for cost parameter used in the model are 4,8,16 and 32. For each cost value, we use L1 and L2 type of loss parameters. The two types of loss functions are:

In L1-SVM, we optimize the following: $minimize \; \frac{1}{2}|w|^2 + C\sum_{i=1}^{M}(\xi_i + \xi^*_i)$

In L2-SVM, we optimize the following: $minimize \; \frac{1}{2}|w|^2 + \frac{C}{2}\sum_{i=1}^{M}(\xi_i + \xi^*_i)^2$

$w$ = regularization term added to avoid overfitting,

$C \in R$ is the penalty parameter. We can tune model performance to ensure a balance between the regularization term and loss function by setting various values of C.

The L1 loss function is called the least absolute deviations (LAD) or error (LAE) and the L2 loss function is also known as the least squares error. In support vector regression, we draw a hyperplane that minimizes the loss function. Hyperplanes change when the loss function changes. We apply the kernel trick to lift the feature space, or convert the lower dimension data into a higher dimension, resulting in a non-linear decision boundary. From the analysis, the model outputs the MAPE in the test set is 9.53% and bias is 3.33% when the cost is 16 and loss is L2. Ridge regression is another model that uses the L2 penalty on the weights of the equation, i.e. it minimizes the L2-regularized squared error instead of only the squared error.

2.2.5. Ridge Regression

When predictors in a regression are strongly correlated, regression coefficients of a variable depends on other predictors in the model. Ridge regression adds bias to alleviate multicollinearity. We fit a model with all $p$ predictors and regularize or constrain the coefficient estimates, i.e. use a method that shrinks the coefficient towards 0 to reduce the variance of the variable. Similar to least square estimates, the ridge regression coefficient estimates $\beta^R$ by minimizing:

$$\sum_{i=1}^{n}(y_i - \beta_0 - \sum_{j=1}^{p}\beta_j x_{ij})^2 + \lambda \sum_{j=1}^{p}\beta_j^2 = RSS + \lambda \sum_{J=1}^{P}\beta_j$$

The penalty term, $\lambda \geq 0$ is the tuning parameter (or the hyperparameter) that is separately determined. $\lambda \sum_{j=1}^{p} \beta_j^2$ is called the shrinkage penalty, which is small when $\beta_1, \beta_2, ...\beta_p$ are close to 0. The shrinkage penalty is not applied to the intercept, $\beta_0$. When $\lambda = 0$, the ridge regression generates least square estimates, but as $\lambda \to \infty$, the effect of the shrinkage penalty increases, which shrinks the coefficients towards 0. In ridge regression, when $\lambda = (-4, 10)$, the test set MAPE is 11% and there is a negative bias of 5.07%. The lowest error rates resulted when the values of hyperparameters were $\alpha = 0$ and $\lambda = 32$. Next, I will discuss the final model incorporated in the ensemble, known as Artificial Neural Networks.

### 2.2.6. Artificial Neural Network

A neural network is a network of interconnected nodes, called neuron. Each neuron is a variable and the intermediary variables are called derived variables. Just as logistic regressions, logit models, models with polynomial transformations and lagged variables are derived from original models and variables, so are the neurons in the hidden layers. All the arrows representing neurons represent parameters, called weights. The output layer contains all the output variables or the output neurons. The layers between input and output layers are called "hidden layers" and they include hidden neurons. The hidden layer is the middle layer which processes information and yields output. In the model, the inputs are $PC_1, PC_2, ..., PC_6$, and the weights are $w_1, w_2, ...w_6$. Thereby, the output is $y = f(x) = \sum PC_i w_i$, where $i = \{1, 2, 3, 4, 5, 6\}$ inputs.

Before building neural network models, I had scaled or normalized data to a standard format to improve the accuracy and speed of training set performance. First, I combined the six principal components with the housing starts variable into a dataframe and then scaled all the explanatory variables: $PC1$ to $PC6$. Since, I have applied the logistical function, I scaled or normalized the the principal components to the interval (0, 1). Then, I further split the dataset into training and test set and created time slices using the same procedure outlined for cross-validation. After pre-processing, I created a neural network model with explanatory variables as $PC1$ to $PC6$ and response variable as housing starts. I added a decay parameter to regularize the network towards reducing bias and variance, so that the models do not overfit the data and generated a neural network of one layer. With the size ranging from 1 to 10, and decay from 0.1 to 4, the the test set MAPE is 10.31% and bias is $-6.63$%.

Figure 19 is a neural interpretation diagram (NID). The black lines indicate positive weights between layers and grey lines indicate negative weights. The thickness of line increases as the magnitude of each weight increases. The first layer consists of input variables whose nodes are labelled as $PC1$ to $PC6$ for the six explanatory variables. It consists of one hidden layer with eight hidden nodes labelled as $H1$ to $H8$, and a bias node labelled as B1. The last layer consists of the output layer labeled as O1. It shows the bias nodes which are connected to the hidden and output layers.

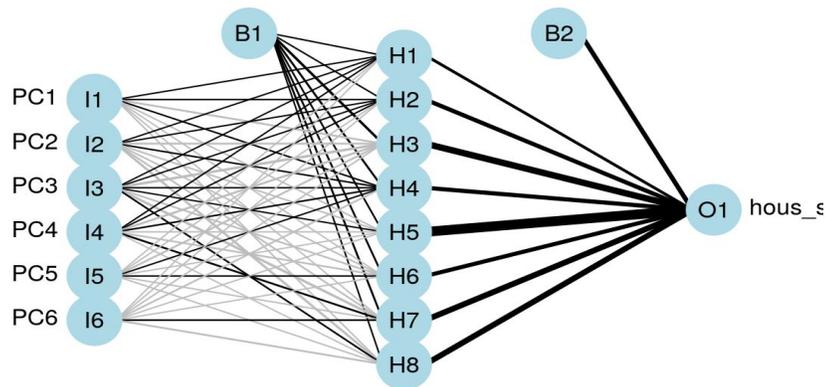

*Figure 14 : Artificial Neural Networks with one hidden layer having eight nodes*

## 2.2.7. Stacked Ensemble

After combining the results from multiple different models to predict housing starts, I created a stacked ensemble which is composed of the K-Nearest Neighbors regression, ridge regression, support vector regression and an artificial neural network. Then, I combined the validation-fold predictions from the component models into a new dataset as explanatory variables. Finally, I stacked a new model via ridge regression where response variable was housing starts and the explanatory variables were the predicted housing starts from each of the component models. Combining predictions from these independent, and less correlated models, reduces variance, and consequently, diminishes the overall test set bias and MAPE.

After obtaining the predictions in the test set and the error rates, I forecasted the values of housing starts from January to December, 2019. We need values of other variables to forecast housing starts in the future, so indirectly I had to predict the principal components as well. We cannot directly use machine learning algorithm for multi-step forecasting. I have made a multivariate multi-step time series forecasting model that forecasts housing over the next twelve months, given the recent and historic level of housing starts and other exogenous variables. So, I have used the ML models recursively to make multi-step forecasts. This procedure predicts one step at a time, feeds the predicted values in the model as inputs to forecast the value for next month.[10] I had iterated this process twelve times to get values from January, 2019 to December, 2019.

## 3. Results

Checking the final models based on the cross-validated performance, I stacked them via ridge regression and obtained the predictions in the test set to compare them with the actual number of housing starts. The test set consists of values from July 2010 to December 2018. The test set MAPE from the stacked ensemble is $6.518\%$ and the bias is $-0.573\%$. The MAPE of the ensemble model is the lowest among the individual component models while the bias is closest to 0, indicating that the ensemble model is unbiased. This implies that on average the predicted value of housing starts deviate from the observed values by 6.518% housing starts per month. Among the econometric models, the $ARIMAX(2, 1, 3)$ has the lowest MAPE and percent. In contrast to the MAPE of the individual learning algorithms and the ensemble model, the econometric methods underperform.

| Econometric and ML Models | MAPE | Percent Bias |
|---|---|---|
| $ARIMA(3, 1, 2)$ | 0.392 | 0.388 |
| $ARIMAX(2, 1, 3)$ | 0.357 | 0.344 |
| $ARMA(2, 2) - GARCH(1, 1)$ | 0.369 | 0.359 |
| $VAR(3)$ | 0.355 | 0.346 |
| KNN | 0.279 | $-0.181$ |
| SVR | 0.095 | 0.033 |
| Ridge Regression | 0.114 | $-0.050$ |
| ANN | 0.103 | $-0.066$ |
| Ensemble of ML models | 0.065 | $-0.005$ |

---

[10] The predictions of each successive month are added at the end of the dataset, which help to make next month's prediction.

*Table 8: Comparing the error rates (MAPE and percent bias) of the econometric, machine learning and ensemble models.*

Figure 20 displays that the predicted and actual housing starts increase marginally over time in the test set.

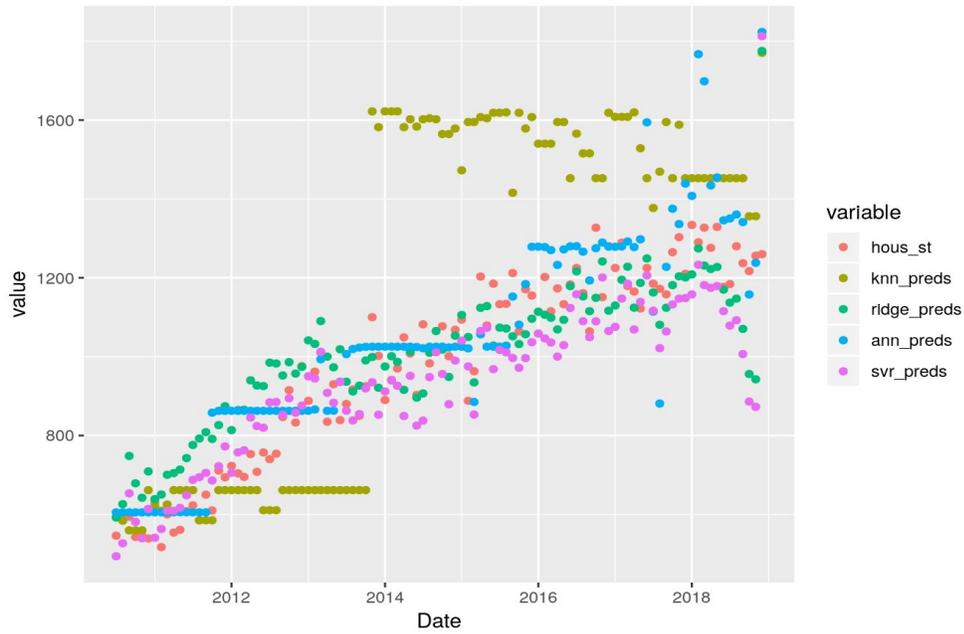

*Figure 15 : Actual and predicted housing starts from each of the component models in the test set*

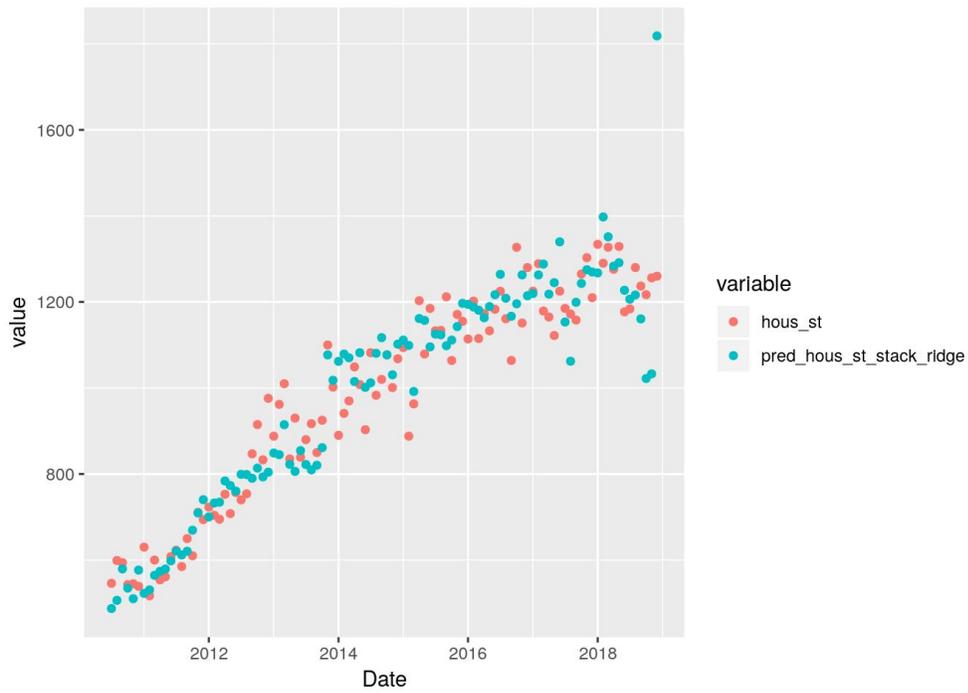

*Figure 16 : Actual Housing starts from July 2010 to December 2018 and predicted Housing starts from the ensemble model*

Based on the increasing order of MAPE, the models perform best in the following order: SVR, ANN, ridge regression, and KNN. This is also corroborated by the plot comparing actual and predicted values from individual models shows − KNN predictions are relatively flat prior to 2014 and substantially lower than the actual value of housing starts. Post-2014, the predictions are exceptionally high, albeit not constant as before. Equivalently, the predictions from ANN are analogous to a step-function: constant before 2012, then jumps up and then flattens between 2012 and late 2013, then surges again twice until it's gradient is positive after year 2015. The predictions from SVR and ridge regression seem to follow the same trend as the actual values of housing starts, but with a few outliers in the years 2017 and 2018.

Finally, Table 9 depicts the forecasts of housing starts (in 1000s of units) from the aforementioned and VARX models for the year 2019. The ARIMA model generates the highest forecasts where the values of housing starts are gradually rising until May, after which it fluctuates. The VARX model gives a more conservative forecast on a lower level, and follows a downward trend throughout the year, widening the gap between the values from VARX and those projected from the other three models. The forecasts from the ARMA−GARCH model are marginally lower than those of the ARIMA model but somewhat linearly climb at a very slow gradient. Finally, the forecasts from surges linearly, then plateaus to some extent, before fluctuating in June-July and then it surges. To sum up, except for VAR, other model forecasts are similar.

| Year 2019 | $ARIMA(3,1,2)$ | $ARIMAX(2,1,3)$ | $VARX(3)$ | $ARMA(2,2)-$ $GARCH(1,1)$ |
|---|---|---|---|---|
| January | 1263.583 | 1180.307 | 1216.2202 | 1255.681 |
| February | 1260.987 | 1225.467 | 1119.0588 | 1258.140 |
| March | 1274.743 | 1227.548 | 1115.0081 | 1259.314 |
| April | 1270.308 | 1228.654 | 1083.1347 | 1260.461 |
| May | 1277.965 | 1224.715 | 1041.3320 | 1261.594 |
| June | 1268.353 | 1222.757 | 1012.1100 | 1262.714 |
| July | 1271.857 | 1182.278 | 982.1425 | 1263.822 |
| August | 1262.992 | 1200.415 | 951.2494 | 1264.91 |
| September | 1268.083 | 1190.270 | 922.9862 | 1265.999 |
| October | 1262.739 | 1195.869 | 896.4428 | 1267.069 |
| November | 1269.046 | 1209.779 | 870.7747 | 1268.127 |
| December | 1264.631 | 1221.963 | 846.4608 | 1269.173 |

*Table 9: Forecasts of housing starts (in 1000s) of units using the four econometric models*

Finally, Table 10 has forecasts from the machine learning models. For all models, except for KNN, the forecasts appear to be cointegrated as they have the same stochastic trends. The lowest forecasts correspond to the SVR model with the lowest MAPE, followed by ridge, ANN and the ensemble model. They follow a downward trend and reach the nadir in October, after which they skyrocket in October-November. The forecasts from KNN are flat till September, then stumble for a month, before spiking and converging with the forecasts from other models. The two plots in the Appendix gives diagrammatically represents the forecasts from both econometric and ML models

| Year 2019 | KNN | Ridge | ANN | SVR | Ensemble |
|---|---|---|---|---|---|
| January | 1452.5 | 1208.5 | 1407.7 | 1157.8 | 1267.4 |
| February | 1452.5 | 1274.4 | 1767 | 1233.1 | 1397.6 |
| March | 1452.5 | 1231 | 1698.3 | 1181.5 | 1351.4 |
| April | 1452.5 | 1222.2 | 1434.3 | 1174.5 | 1283.3 |
| May | 1452.5 | 1227.5 | 1454.4 | 1179 | 1291.2 |
| June | 1452.5 | 1170 | 1345.7 | 1115.3 | 1227.2 |
| July | 1452.5 | 1137 | 1350.4 | 1078.7 | 1206.5 |
| August | 1452.5 | 1147.2 | 1360.4 | 1092.2 | 1216.3 |
| September | 1452.5 | 1070.4 | 1341.2 | 1006.5 | 1160.7 |
| October | 1356 | 956.1 | 1157.6 | 886 | 1022.1 |
| November | 1356 | 942.6 | 1238 | 872.7 | 1032.8 |
| December | 1770.5 | 1775.1 | 1823.3 | 1812.8 | 1818.4 |

*Table 10: Forecasts of housing starts (in 1000s) of units using the four ML and ensemble models*

## 4. Conclusion

In this empirical analysis, I have conducted time series analysis and forecasted housing starts using various econometric and machine learning models. From the results, I observed that the ML models outperform the econometric models. The ensemble ML model had a MAPE of 6.51 percent compared to the MAPE of $ARIMAX(2,1,3)$ which was 35.7 percent − lowest among all other econometric models. This suggests that economists engaged in forecasting macroeconomic variables should explore forecasting from ML based models. These models can discover complex non-linear relationships in the data, without assuming anything about the exogenous factors. They can determine the relative importance of each variable without being affected by multicollinearity. However, we also have to be prudent when deciding which models to use, particularly, as they cannot be interpreted and can overfit in the training sets.

In both time series and ML models, performance of forecasts differ depending on the dataset. As we often assume that past patterns can indicate the behavior of a series, they are projected. Consequently, if the patterns continue, the forecasts will be precise, but if the patterns abruptly deflect, the projections may heavily differ from the actual value, as noticed from the flat KNN predictions. This creates a "black swan" event wherein the event deviates beyond its generally expected path and is hard to project. So, we have to retrain the model repeatedly to account for newer information. ML models can capably model any type of patterns, compared to time series forecasting methods −where we have first ensure homoskedastic errors and same probability distribution throughout the series. Hence, I have conducted several tests such as (G)ARCH for conditional heteroskedasticity, and Johansen test of cointegration, among others. Thereafter, I have forecasted housing starts.

# 5. Appendix

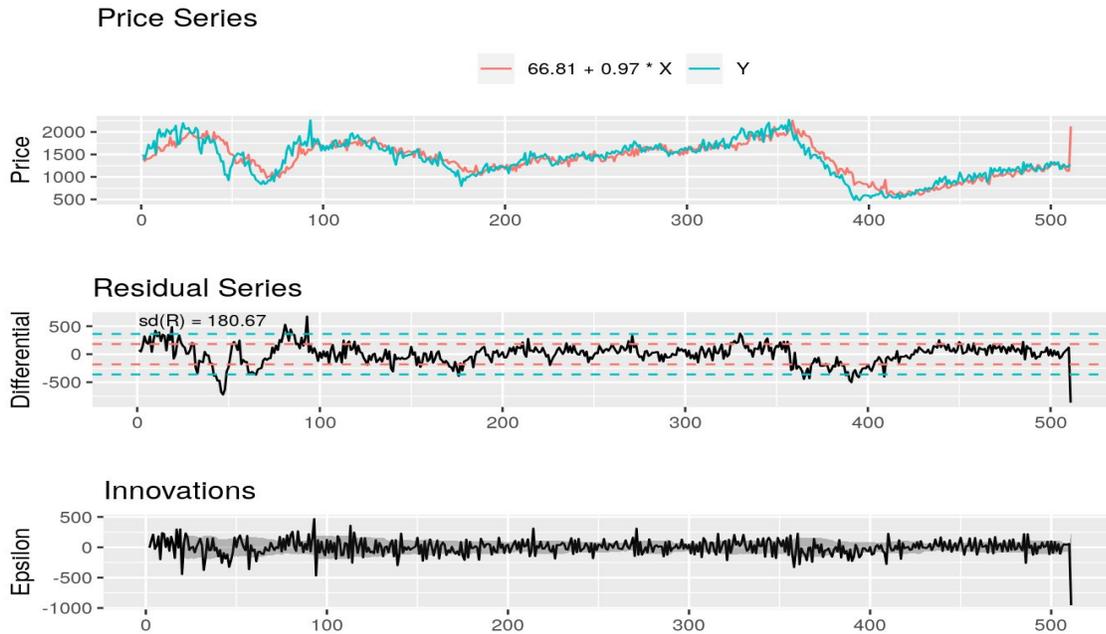

*Figure 1: Plots showing the equation, residuals and innovations from the cointegration model between housing starts and private houses completed*

| hous_st | hous_st | pvt_house_comp | mortgR | house_supply |
| --- | --- | --- | --- | --- |
| ECT1 | -0.0268(0.0425) | 0.3199(0.0347) | 0.0003(0.0001) | 0.0003(0.0002) |
| ECT2 | -0.0363(0.0375) | -0.3288(0.0307) | -0.0002(9.8e-05) | -0.0002(0.0002) |
| Intercept | 163.2674(34.76) | -9.3985(28.4199) | -0.2161(0.0905) | -0.2639(0.1756) |
| $hous\ st_{t-1}$ | -0.5518(0.0587) | -0.2430(0.0480) | 1.7e-06(0.0002) | -0.0003(0.0003) |
| $hous\ st_{t-2}$ | -0.3224(0.0576) | -0.2111(0.0471) | -0.0001(0.0001) | 1.2e-05(0.0003) |
| $hous\ st_{t-3}$ | -0.1176(0.0489) | -0.1207(0.0400) | -0.0002(0.0001) | -5.1e-05(0.0002) |
| $mortgR_{t-1}$ | -43.9864(18.0267) | 2.5715(14.7387) | 0.5653(0.0469) | 0.6693(0.0911) |
| $mortgR_{t-2}$ | -11.1202(19.6690) | 7.4237(16.0815) | -0.3211(0.0512) | -0.0641(0.0993) |
| $mortgR_{t-3}$ | -42.0914(18.5365) | -1.7610(15.1556) | 0.1238(0.0482) | 0.1072(0.0936) |
| $pvt\ hous\ comp_{t-1}$ | 0.0496(0.0596) | -0.4583(0.0487) | 0.0002(0.0002) | 0.0008(0.0003) |
| $pvt\ hous\ comp_{t-2}$ | -0.0044(0.0663) | -0.2872(0.0542) | 0.0002(0.0002) | 0.0007(0.0003) |
| $pvt\ hous\ comp_{t-3}$ | -0.0131(0.0597) | -0.1894(0.0488) | 0.0001(0.0002) | 0.0002(0.0003) |
| $hous\ supply_{t-1}$ | -13.1016(10.4502) | -4.5648(8.5442) | -0.0506(0.0272) | -0.3478(0.0528) |
| $hous\ supply_{t-2}$ | -21.3106(10.5554) | -5.8704(8.6302) | -0.0625(0.0275) | -0.2075(0.0533) |
| $hous\ supply_{t-3}$ | -14.1123(9.9092) | -4.0074(8.1018) | -0.0483(0.0258) | 0.0047(0.0501) |

*Table 1: Coefficients of the VECM*

```
================================================================
                               Dependent variable:
                    --------------------------------------------
                    hous_st, pvt_house_comp, mortgR, house_supply
                       (1)        (2)         (3)         (4)
----------------------------------------------------------------
hous_st.l1           0.443***   0.097**    0.0002**    0.00003
                     (0.046)    (0.038)    (0.0001)   (0.0002)

pvt_house_comp.l1     0.040    0.283***    0.0001      0.001*
                     (0.064)    (0.053)    (0.0002)   (0.0003)

mortgR.l1           -32.268*    0.294      1.512***   0.639***
                    (17.514)   (14.556)    (0.045)    (0.087)
```

| | | | | |
|---|---:|---:|---:|---:|
| house_supply.l1 | -29.253*** | -3.226 | -0.018 | 0.661*** |
| | (9.570) | (7.954) | (0.025) | (0.047) |
| hous_st.l2 | 0.222*** | 0.044 | -0.0001 | 0.0003 |
| | (0.050) | (0.041) | (0.0001) | (0.0002) |
| pvt_house_comp.l2 | -0.039 | 0.250*** | 0.00001 | -0.0002 |
| | (0.063) | (0.053) | (0.0002) | (0.0003) |
| mortgR.l2 | -0.614 | 6.542 | -0.802*** | -0.639*** |
| | (28.479) | (23.670) | (0.074) | (0.141) |
| house_supply.l2 | -7.547 | 1.884 | -0.003 | 0.139** |
| | (11.436) | (9.505) | (0.030) | (0.056) |
| hous_st.l3 | 0.227*** | 0.137*** | -0.0001 | -0.0001 |
| | (0.048) | (0.040) | (0.0001) | (0.0002) |
| pvt_house_comp.l3 | 0.025 | 0.183*** | -0.0001 | -0.0005 |
| | (0.060) | (0.050) | (0.0002) | (0.0003) |
| mortgR.l3 | 30.923* | -5.984 | 0.274*** | 0.013 |
| | (17.955) | (14.923) | (0.047) | (0.089) |
| house_supply.l3 | 13.121 | 5.803 | 0.032 | 0.215*** |
| | (9.800) | (8.146) | (0.025) | (0.048) |
| const | 321.561*** | -41.733 | 0.022 | -0.805*** |
| | (62.661) | (52.080) | (0.162) | (0.309) |
| trend | -0.193*** | 0.034 | -0.0003 | 0.001** |
| | (0.070) | (0.058) | (0.0002) | (0.0003) |

```
-------------------------------------------------------------------------
Observations                       508         508         508         508
R2                                 0.944       0.954       0.995       0.922
Adjusted R2                        0.943       0.953       0.994       0.919
Residual Std. Error (df = 494)     97.370      80.928      0.252       0.481
F Statistic (df = 13; 494)         642.237***  794.994***  6,916.869*** 446.179***
=========================================================================
Note:                                              *p<0.1; **p<0.05; ***p<0.01
```

*Table 2: R output: VARX(3) model*

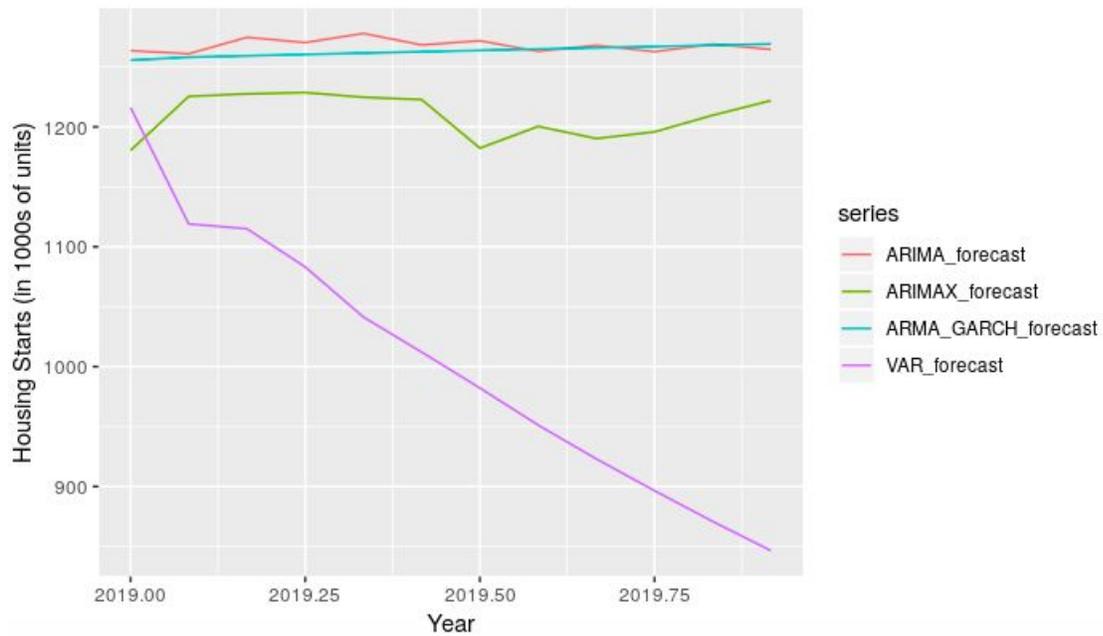

*Figure 2: The forecasts from ARIMA, ARIMAX, VARX and ARMA - GARCH models for year 2019*

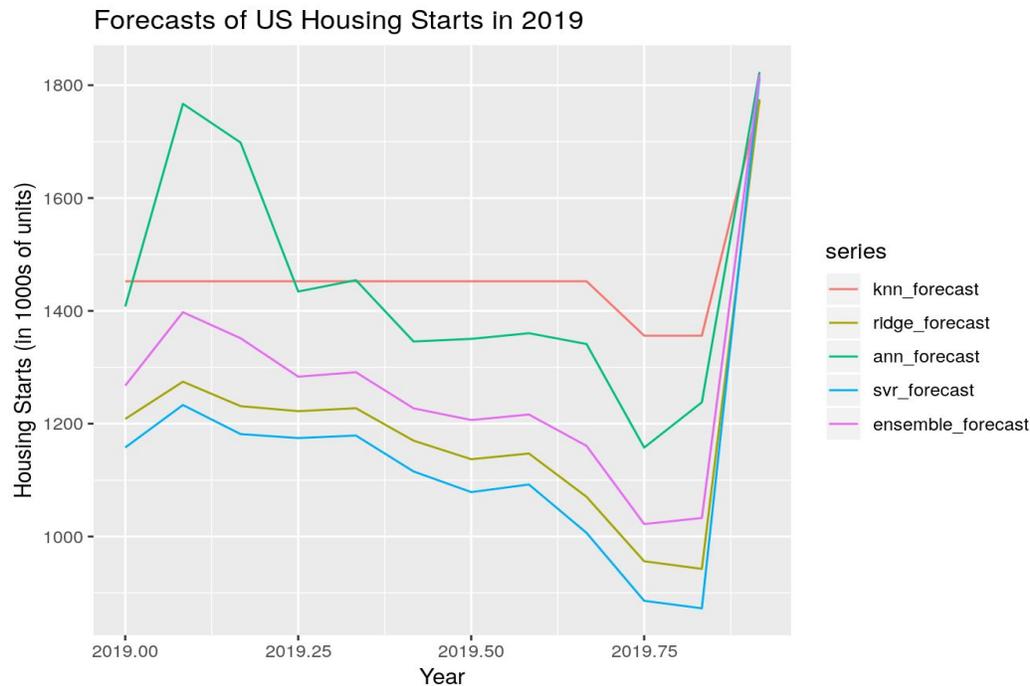

*Figure 3: The forecasts from KNN, ANN, Ridge, SVR and ensemble models for year 2019*

## **6. Bibliography**

1. Brady, Eugene A. "Forecasting Housing Starts." *Business Economics* (1971): 18-22.
2. Coccari, Ronald L. "Time series analysis of new private housing starts." *Business Economics* (1979): 95-109.
3. Hansen, James V., James B. McDonald, and Ray D. Nelson. "Some evidence on forecasting time-series with Support Vector Machines." *Journal of the Operational Research Society* 57, no. 9 (2006): 1053-1063.
4. Henly, Samuel, and Alexander Wolman. "Housing and the Great Recession: a VAR accounting exercise." (2011).
5. Joseph, Anthony, and Maurice Larrain. "Housing Starts Forecast of Retail Sales through the 2007-2009 Recession." *Procedia Computer Science* 12 (2012): 271-275.
6. Khalafallah, Ahmed. "Neural network based model for predicting housing market performance." Tsinghua Science and Technology 13, no. S1 (2008): 325-328.

7. Kemme, David M., and Saktinil Roy. "Did the recent housing boom signal the global financial crisis?." *Southern Economic Journal* 78, no. 3 (2012): 999-1018.

8. Kawaller, Ira G., and Timothy W. Koch. "Housing as a Monetary Phenomenon: Forecasting Housing Starts Using the Monetary Aggregate Targets." *Business Economics* (1981): 30-35.

9. Qiu, Xueheng, Le Zhang, Ye Ren, Ponnuthurai N. Suganthan, and Gehan Amaratunga. "Ensemble deep learning for regression and time series forecasting." In *Computational Intelligence in Ensemble Learning (CIEL), 2014 IEEE Symposium on*, pp. 1-6. IEEE, 2014.

10. Gu, Jirong, Mingcang Zhu, and Liuguangyan Jiang. "Housing price forecasting based on genetic algorithm and support vector machine." *Expert Systems with Applications* 38, no. 4 (2011): 3383-3386.

11. Dua, Pami, Stephen M. Miller, and David J. Smyth. "Using leading indicators to forecast US home sales in a Bayesian vector autoregressive framework." *The Journal of Real Estate Finance and Economics* 18, no. 2 (1999): 191-205.

12. Hsu, Ming-Wei, Stefan Lessmann, Ming-Chien Sung, Tiejun Ma, and Johnnie EV Johnson. "Bridging the divide in financial market forecasting: machine learners vs. financial economists." *Expert Systems with Applications* 61 (2016): 215-234.

13. Pavlyshenko, Bohdan M. "Machine-Learning Models for Sales Time Series Forecasting." *Data* 4, no. 1 (2019): 15.

14. Makridakis, Spyros, Evangelos Spiliotis, and Vassilios Assimakopoulos. "Statistical and Machine Learning forecasting methods: Concerns and ways forward." *PloS one* 13, no. 3 (2018): e0194889.

15. Engle, Robert. "GARCH 101: The use of ARCH/GARCH models in applied econometrics." *Journal of economic perspectives* 15, no. 4 (2001): 157-168.

16. Park, Byeonghwa, and Jae Kwon Bae. "Using machine learning algorithms for housing price prediction: The case of Fairfax County, Virginia housing data." *Expert Systems with Applications* 42, no. 6 (2015): 2928-2934.

17. Ahmed, Nesreen K., Amir F. Atiya, Neamat El Gayar, and Hisham El-Shishiny. "An empirical comparison of machine learning models for time series forecasting." *Econometric Reviews* 29, no. 5-6 (2010): 594-621.

18. Jung, Jin-Kyu, Manasa Patnam, and Anna Ter-Martirosyan. *An Algorithmic Crystal Ball:Forecasts-based on Machine Learning*. International Monetary Fund, 2018.

19. Luo, H. and Wang, S., 2017. Based on the PCA-ARIMA-BP hybrid model of stock price prediction research. *ANZIAM Journal*, *58*, pp.162-178.

20. Bergmeir, C., Hyndman, R.J. and Koo, B., 2018. A note on the validity of cross-validation for evaluating autoregressive time series prediction. *Computational Statistics & Data Analysis*, *120*, pp.70-83.

21. James, Gareth, Daniela Witten, Trevor J. Hastie, and Robert J. Tibshirani. *An Introduction to Statistical Learning: With Applications in R*. New York: Springer, 2017.

22. "Federal Reserve Economic Data | FRED | St. Louis Fed." FRED. Accessed May 01, 2019. https://fred.stlouisfed.org/.